\pdfoutput=1
\documentclass[11pt,twoside,a4paper,cmspaper,final,collab]{cms-tdr}
\def\svnVersion{f85390c}\def\svnDate{2025/02/09}\def\cmsCernNoTag{CERN-EP-2024-266}\def\cmsCernDate{\today}\def\cmsMessage{Published in the European Physical Journal C as \href{http://dx.doi.org/10.1140/epjc/s10052-024-13729-y}{\doi{10.1140/epjc/s10052-024-13729-y}.}}
\begin{document}\cmsNoteHeader{B2G-22-005}

\newlength\cmsFigWidth
\ifthenelse{\boolean{cms@external}}{\setlength\cmsFigWidth{0.45\textwidth}}{\setlength\cmsFigWidth{0.49\textwidth}}
\ifthenelse{\boolean{cms@external}}{\providecommand{\cmsLeft}{upper\xspace}}{\providecommand{\cmsLeft}{left\xspace}}
\ifthenelse{\boolean{cms@external}}{\providecommand{\cmsRight}{lower\xspace}}{\providecommand{\cmsRight}{right\xspace}}
\newlength\cmsTabSkip\setlength{\cmsTabSkip}{2ex}

\newcommand{\tstar}{{\HepParticle{\PQt}{}{\ast}}\xspace}
\newcommand{\tstartstar}{\ensuremath{\tstar\HepAntiParticle{\PQt}{}{*}}\xspace}
\newcommand{\tstartg}{\ensuremath{\tstar \to \PQt\Pg}\xspace}
\newcommand{\tstarta}{\ensuremath{\tstar \to \PQt\PGg}\xspace}
\newcommand{\tstartstartgtg}{\ensuremath{\tstartstar \to \PQt\Pg\PAQt\Pg}\xspace}
\newcommand{\mtstar}{\ensuremath{m_{\tstar}}\xspace}
\DeclareRobustCommand{\PV}{{\HepParticle{V}{}{}}\Xspace}
\newcommand{\ST}{\ensuremath{S_{\text{T}}}\xspace}
\newcommand{\pp}{\ensuremath{\Pp\Pp}\xspace}
\newcommand{\Irel}{\ensuremath{I_{\text{rel}}}\xspace}
\newcommand{\BR}{\ensuremath{\mathcal{B}}\xspace}
\newcommand{\sDNN}{\ensuremath{s_{\text{DNN}}}\xspace}
\newcommand{\sDDT}{\ensuremath{s_{\text{DDT}}}\xspace}
\newcommand{\gTF}{\ensuremath{g_{\text{TF}}}\xspace}
\newcommand{\eptt}{\ensuremath{\epsilon_{\ttbar}}\xspace}

\cmsNoteHeader{B2G-22-005}
\title{Search for pair production of heavy particles decaying to a top quark and a gluon in the lepton+jets final state in proton-proton collisions at \texorpdfstring{$\sqrt{s}=13\TeV$}{sqrt(s)=13 TeV}}
\titlerunning{Search for \tstartstartgtg in the lepton+jets final state}

\author*{The CMS Collaboration}

\date{\today}

\abstract{
A search is presented for the pair production of new heavy resonances, each decaying into a top quark (\PQt) or antiquark and a gluon (\Pg). The analysis uses data recorded with the CMS detector from proton-proton collisions at a center-of-mass energy of 13\TeV at the LHC, corresponding to an integrated luminosity of 138\fbinv. Events with one muon or electron, multiple jets, and missing transverse momentum are selected. After using a deep neural network to enrich the data sample with signal-like events, distributions in the scalar sum of the transverse momenta of all reconstructed objects are analyzed in the search for a signal. No significant deviations from the standard model prediction are found. Upper limits at 95\% confidence level are set on the product of cross section and branching fraction squared for the pair production of excited top quarks in the \tstartg decay channel. The upper limits range from 120 to 0.8\unit{fb} for a \tstar with spin-1/2 and from 15 to 1.0\unit{fb} for a \tstar with spin-3/2. These correspond to mass exclusion limits up to 1050 and 1700\GeV for spin-1/2 and spin-3/2 \tstar particles, respectively. These are the most stringent limits to date on the existence of \tstartg resonances.
}

\hypersetup{
pdfauthor={CMS Collaboration},
pdftitle={Search for pair production of heavy particles decaying to a top quark and a gluon in the lepton+jets final state in proton-proton collisions at sqrt(s)=13 TeV},
pdfsubject={CMS},
pdfkeywords={CMS, excited top quark, vector-like quark, B2G}}

\maketitle 

\section{Introduction}
\label{sec:introduction}

{\tolerance=800
The standard model (SM) of particle physics has proven to be a highly accurate and successful theory. 
It describes the properties of all elementary particles and their interactions with a
small number of free parameters. 
However, several observations indicate that it is not a complete theory and has to be extended. 
For example, the naturalness problem arises because of quadratically divergent contributions to the Higgs boson mass 
from the top quark and the \PW and \PZ bosons. In the SM, an unnaturally large fine-tuning is required to explain the 
measured Higgs boson mass of around 125\GeV~\cite{higgsdiscatlas, higgsdisccms, higgsdisccms2}. 
This motivates searches for possible extensions of the SM, particularly for theoretical models proposing that the top quark is not a fundamental 
but a composite particle~\cite{composite1, composite2, composite3, composite4, composite5, composite6, composite8}. 
A composite top quark would result in excited states that could be observed at the CERN LHC, 
providing a way to confirm or exclude these models. 
Excited top quarks \tstar are predicted to decay instantly after their production into 
a top quark by the radiation of excess energy in the form of a gluon or a photon, \tstartg or \tstarta. 
Other models of physics beyond the SM propose the existence of vector-like fermionic top quark partners, 
for example little Higgs models~\cite{littleHiggs1, littleHiggs2, littleHiggs3, littleHiggs4}, 
models with extra dimensions~\cite{extradim1, extradim2, extradim3}, or composite Higgs models~\cite{compositehiggs1}. 
Usually, searches for these types of models are performed targeting decays including \PW, \PZ, or Higgs 
bosons~\cite{CMS_bW_pair, CMS_tH_tZ_single, CMS_tZ_single, CMS_tH_tZ_pair, ATLAS_tH_tZ_bW_pair, ATLAS_tZ_pair, ATLAS_tZ_single}. 
However, in the case of a small mixing between the top quark partner and the top quark, these decay modes are suppressed 
and the quantum loop induced decays \tstartg and \tstarta become dominant~\cite{Alhazmi:2018whk}. 
Top quark partners can provide an elegant solution to the naturalness problem, 
introducing additional terms cancelling out the quadratic divergences~\cite{toploop}.
\par}

In this article, we present a search for the pair production of heavy excited top quarks or top quark partners, 
generically labeled \tstar in the following. We target the \tstartstartgtg channel with an experimental 
signature that is independent of which underlying theory is predicting it. The \tstar is predicted to carry the same weak isospin, 
color and weak hypercharge as the top quark, but could exist with different spins: spin-1/2 or spin-3/2~\cite{Alhazmi:2018whk}. 
For a spin-1/2 \tstar, the field can be described in a similar way to those of heavy SM quarks.
The spin-3/2 case is described by a Rarita--Schwinger Lagrangian~\cite{rarita_schwinger}, 
adapting the Dirac equation to a spin-3/2 particle. Assuming equal couplings to the strong and electroweak sectors 
compared with the SM top quark leads to a branching fraction \BR of 97\% for the decay \tstartg 
in both spin scenarios~\cite{Alhazmi:2018whk}. This motivates a search where both \tstar decay 
into a top quark and a gluon. 

\begin{figure}[t]
    \centering
    \includegraphics[width=\cmsFigWidth]{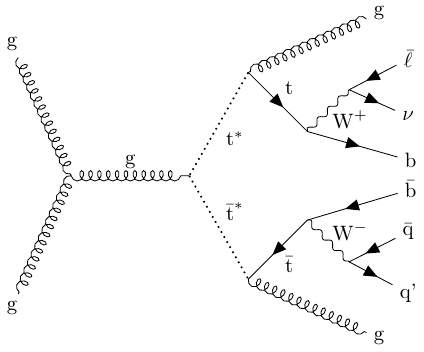}
    \caption{Representative Feynman diagram of the signal process at leading order.}
    \label{fig:feynman}
\end{figure}

We search for the \tstartstartgtg process in final states containing a single lepton, a neutrino, 
four light jets from two gluons and two light quarks, and two jets originating from \PQb quarks. 
A representative Feynman diagram of a signal process is shown in Fig.~\ref{fig:feynman}. 
Data collected between 2016 and 2018 with the CMS detector in proton-proton (\pp) collisions at a 
center-of-mass energy of 13\TeV at the LHC are used in this analysis, corresponding to an integrated 
luminosity of 138\fbinv. Previous searches in the same final state have been performed by the CMS Collaboration 
using 8\TeV~\cite{CMS:2013tlq} and 13\TeV~\cite{CMS:2017ixp} \pp data. In comparison to the previous 13\TeV result, 
which is based on data corresponding to an integrated luminosity of 35.9\fbinv, the analysis presented in this article 
achieves a substantial increase in sensitivity from the larger data set and 
from improved analysis techniques. In particular, machine-learning techniques are used to suppress backgrounds 
from known SM processes, and a more inclusive observable for the signal extraction 
results in an improved signal efficiency compared to the previous 13\TeV analysis~\cite{CMS:2017ixp}. 

{\tolerance=800
Tabulated results are provided in the HEPData record for this analysis~\cite{hepdata}.
\par}

\section{The CMS detector and simulated samples}
\label{sec:samples}

The central feature of the CMS apparatus is a superconducting solenoid of 6\unit{m} internal diameter, 
providing a magnetic field of 3.8\unit{T}. Within the solenoid volume are a silicon pixel and strip tracker, 
a lead tungstate crystal electromagnetic calorimeter (ECAL), and a brass and scintillator hadron calorimeter (HCAL), 
each composed of a barrel and two endcap sections. Forward calorimeters extend the pseudorapidity ($\eta$) coverage provided 
by the barrel and endcap detectors. Muons are measured in gas-ionization detectors embedded in the steel flux-return 
yoke outside the solenoid. More detailed descriptions of the CMS detector, together with a definition of the 
coordinate system used and the relevant kinematic variables, can be found in Refs.~\cite{CMS:2008xjf,CMS:2023gfb}. 

Events of interest are selected using a two-tiered trigger system. The first level, 
composed of custom hardware processors, uses information from the calorimeters and muon detectors to select 
events at a rate of around 100\unit{kHz} within a fixed latency of 4\mus~\cite{CMS:2020cmk}. 
The second level, known as the high-level trigger (HLT), consists of a farm of processors running a version of the 
full event reconstruction software optimized for fast processing, and reduces the event rate to 
around 1\unit{kHz} before data storage~\cite{CMS:2016ngn}.

Considering both the spin-1/2 and spin-3/2 scenarios, we produce Monte Carlo (MC) simulated samples of the 
$\pp \to \tstartstar$ signal process at leading order (LO) in perturbative quantum chromodynamics (QCD), 
using \MGvATNLO~2.6.5~\cite{Alwall:2014hca}. We consider different values of resonance masses \mtstar 
between 700 and 3000\GeV, in steps of 100\GeV below 2000\GeV and in steps of 250\GeV above. 
The interactions with the SM particles are simulated using an effective field theory (EFT)~\cite{Dicus:2012uh, Alhazmi:2018whk}, 
with the assumption that the \tstar couplings in the strong and electroweak sectors are identical to the top quark couplings. 
The predicted production cross sections for the pair production of spin-3/2 \tstar particles is larger than 
for spin-1/2 particles of the same mass, as the cross section for spin-3/2 grows with energy as $\hat{s}^3$~\cite{Dicus:2012uh}, 
where $\hat{s}$ is the partonic center-of-mass energy squared. This violates unitarity at high energies,
but is regulated by some large scale of new physics $\Lambda \gg \mtstar$ at which the EFT approach becomes invalid. 
Differences in the kinematic distributions of \tstar systems with different spins are observed in angular correlations, 
as well as in the \tstar momenta and the invariant mass of the \tstartstar system, where 
the momenta and mass are larger on average for spin-3/2 compared to spin-1/2 fermions. 
These spin-dependent differences become smaller with increasing \mtstar.

{\tolerance=800
Relevant SM background processes for this search are top quark pair (\ttbar) and single top quark production,
as well as the production of a \PW boson in association with jets (\PW{}+jets).
Less important background processes include the production of two weak gauge bosons (diboson),
the production of multiple jets from the strong interaction (QCD multijet) and the Drell--Yan process.
The \ttbar production is simulated at next-to-LO (NLO) with 
\POWHEG~v2~\cite{Nason:2004rx,Frixione:2007vw,Frixione:2007nw,Alioli:2010xd,Re:2010bp}. 
The cross section of the \ttbar process is corrected to a prediction at next-to-NLO (NNLO) precision, 
using a next-to-next-to-leading-logarithmic soft-gluon approximation, obtained with the \textsc{Top++}~2.0 program~\cite{ttnnlo}. 
The production of single top quarks is simulated at NLO using \POWHEG for the $\PQt\PW$ and $t$ channels, 
and \MGvATNLO for the $s$ channel. The \PW{}+jets production is 
simulated at LO with four additional partons with \MGvATNLO. Diboson, QCD multijet 
production, and the Drell--Yan process are simulated at LO with \PYTHIA~8.240~\cite{Sjostrand:2014zea}.
For all generated samples, the NNPDF~3.1~\cite{NNPDF} NNLO parton distribution function (PDF) sets are used. 
\par}

{\tolerance=800
For all generated collision events, parton shower and hadronization are simulated using \PYTHIA with 
the CP5 tune~\cite{CMS:2015wcf}. The CMS detector response is simulated using \GEANTfour~\cite{GEANT4:2002zbu}. 
Additional inelastic \pp collision events are simulated using \PYTHIA and superimposed on simulated events 
to model the effect of additional \pp collisions within the same or adjacent bunch crossings (pileup). 
We use a total inelastic cross section of 69.2\unit{mb}~\cite{CMS:2018mlc} to estimate the expected number 
of \pp interactions per bunch crossing and correct the simulation to match the corresponding 
distribution to that observed in data.
\par}

\section{Event reconstruction}
\label{sec:eventreconstruction}

A particle-flow (PF) algorithm~\cite{CMS:2017yfk} aims to reconstruct and identify each particle
in an event, with an optimized combination of information from the various elements of the CMS detector.

The energy of muons is obtained from the curvature of the corresponding track. These are measured with a 
reconstruction and identification efficiency greater than 96\%. For muons with transverse momentum \pt up to 100\GeV, 
a relative transverse momentum resolution of 1\% in the barrel and 3\% in the endcaps is obtained. 
The \pt resolution in the barrel is better than 7\% for muons with \pt up to 1\TeV~\cite{CMS:2018rym}. 
A set of selection criteria, different for low-\pt~\cite{muID_cutbased_2} and high-\pt~\cite{muID_highpt} muons, 
is used to select prompt muons. Scale factors are used to correct for observed
differences between data and simulation in the muon reconstruction and selection efficiencies.

The energy of electrons is determined from a combination of the electron momentum at the primary interaction 
vertex as determined by the tracker, the energy of the corresponding ECAL cluster, and the energy sum of all 
bremsstrahlung photons spatially compatible with originating from the electron track~\cite{CMS:2020uim}.  
The momentum resolution for electrons with $\pt \approx 45\GeV$ from $\PZ \to \Pe \Pe$ decays ranges from 1.6 to 5\%. 
It is generally better in the barrel region than in the endcaps, and also depends on the bremsstrahlung energy 
emitted by the electron as it traverses the material in front of the ECAL~\cite{CMS:2020uim,CMS-DP-2020-021}. 
A multivariate selection with a 90\% efficiency to identify prompt electrons is used. 
Correction scale factors ensure that the electron reconstruction and selection efficiency is well modeled in MC simulation.

For both muons and electrons, we define an isolation variable \Irel as the \pt sum of nearby PF candidates 
relative to the lepton \pt, after accounting for pileup contributions~\cite{CMS:2020uim, CMS:2018rym}. 
For the calculation of \Irel, PF candidates within $\DR < 0.4$ and 0.3 for muons and electrons, respectively, are considered. 
The angular distance \DR between two PF candidates $i$ and $j$ is defined as $\DR = \sqrt{\smash[b]{ (\Delta \eta_{i,j})^2 + (\Delta \phi_{i,j})^2 }}$, 
where $\Delta \eta_{i,j}$ and $\Delta \phi_{i,j}$ denote the distances in pseudorapidity and azimuth, respectively. 

The energy of photons is obtained from the ECAL measurement~\cite{CMS:2020uim}. 
The energy of charged hadrons is determined from a combination of their momentum measured in the tracker and the 
matching ECAL and HCAL energy deposits, corrected for the response function of the calorimeters to hadronic showers. 
The energy of neutral hadrons is obtained from the corresponding corrected ECAL and HCAL energies. 
Two different types of hadron jets are reconstructed from PF candidates, referred to as small-radius and variable-radius jets.

{\tolerance=800
Small-radius jets are clustered using the anti-\kt algorithm~\cite{Cacciari:2008gp} as implemented in 
the \FASTJET package~\cite{Cacciari:2011ma} with an angular distance parameter of 0.4. 
Jet momentum is determined as the vectorial sum of all particle momenta in the jet, and is found from simulation to be, 
on average, within 5 to 10\% of the true momentum over the entire \pt spectrum and detector acceptance. 
Pileup interactions can contribute additional tracks and calorimetric energy depositions, 
increasing the apparent jet momentum. The pileup-per-particle identification (PUPPI) algorithm~\cite{Bertolini:2014bba,CMS:2020ebo} 
is used to mitigate the effect of pileup at the reconstructed-particle level, using local shape information, 
event pileup properties, and tracking information. Jet energy corrections are derived from simulation studies so 
that the average measured energy of jets becomes identical to the truth information from the simulation. 
In situ measurements of the momentum balance in dijet, $\PGg$+jet, \PZ{}+jet, and multijet events are used to 
determine any residual differences between the jet energy scale in data and in simulation, 
and appropriate corrections are made~\cite{CMS:2016lmd}. Additional selection criteria are applied to each 
jet to remove jets potentially dominated by instrumental effects or reconstruction failures~\cite{CMS:2020ebo}. 
\par}

{\tolerance=800
The second jet reconstruction algorithm, used to reconstruct variable-radius jets,
is the Heavy Object Tagger with Variable $R$ (HOTVR) algorithm~\cite{Lapsien:2016zor}, 
which is also implemented in \FASTJET. It combines jet clustering with subjet finding and a soft-cluster rejection. 
Resulting jets have an angular size between 0.1 and 1.5 in the jet distance parameter, 
where the active jet area~\cite{Cacciari:2008gn} decreases with increasing momentum. 
This jet reconstruction algorithm has been developed for the reconstruction of 
fully hadronic decays of Lorentz-boosted top quarks, where each decay results in a single jet reconstructed with an optimal jet size~\cite{CMS:2020poo}. 
The top quark momentum, and therefore its Lorentz boost, varies depending on \mtstar in this search.
Variable-radius jets are thus better suited to capture the kinematic properties of the top quarks with reduced 
dependence on the assumed signal mass~\cite{Kogler:2021kkw}. 
Jets originating from gluons of the \tstar decay are reconstructed with the HOTVR algorithm as well, where the variable size 
of these jets helps to capture wide-angle radiation, which would result in out-of-cone effects for small-radius jets. 
The substructure of variable-radius jets is analyzed using the $N$-subjettiness variables $\tau_N$, 
which provide a measure of the degree to which jets contain $N$ or fewer localized regions of high energy density~\cite{taun, Thaler:2011gf}. 
Pileup is accounted for using the PUPPI algorithm, following a similar procedure to that used for the small-radius jets.
Momentum corrections are applied to the individual subjets of each HOTVR jet. 
Recombination of the corrected subjets gives the corrected variable-radius jet. 
This procedure was first applied in Ref.~\cite{CMS:2021mku} and has been validated again for this analysis. 
\par}

{\tolerance=1200
Jets originating from the fragmentation of \PQb quarks are identified using the multiclassification algorithm 
\textsc{DeepJet}~\cite{BTV-16-002, Bols:2020bkb, CMS-DP-2023-005}, based on a deep neural network (DNN). 
It uses information from jet constituents, charged and neutral particles, secondary vertices, and global event 
variables to define a score for each jet. This score quantifies how likely the jet is to have originated from the decay of a \PQb quark. 
Small-radius with a \textsc{DeepJet} score above a certain threshold, corresponding to a 1\% misidentification probability of light-quark and gluon jets, 
are referred to as \PQb jets. This working point has an efficiency between 70 and 80\% for correctly identifying \PQb jets, depending on the data-taking era.
Differences between data and simulation in the shape of the \textsc{DeepJet} score distribution are accounted for by applying correction factors, 
and a secondary correction ensures that no change in normalization is introduced.
\par}

The missing transverse momentum vector \ptvecmiss is computed as the negative vector sum of the transverse momenta 
of all PF candidates in an event, and its magnitude is denoted as \ptmiss~\cite{CMS:2019ctu}. 
The PUPPI algorithm is applied to reduce the pileup dependence of the \ptvecmiss observable. 
The \ptvecmiss is modified to account for corrections to the energy scale of the jets in the event.

\section{Event selection}
\label{sec:selection}

We use HLT algorithms requiring the presence of a single muon, electron, or photon. 
These select events with at least one muon having \pt above 50\GeV or an electron with \pt above 115\GeV, 
with no isolation requirements. Complementary triggers requiring an isolated lepton extend the reach of 
the analysis to lower \pt, requiring a muon with \pt of at least 24 or 27\GeV in 2016 or 2017--2018,
or an electron with \pt of at least 27, 35, or 32\GeV in the years 2016, 2017, and 2018, respectively. 
The different \pt thresholds were introduced because of different instantaneous luminosities during the three years of data taking. 
To recover inefficiencies at high electron \pt, events selected using photon triggers with a photon with \pt larger than 
175 or 200\GeV in 2016 or 2017--2018, respectively, are considered as well. The muon trigger efficiency exceeds 90\% 
over the full momentum range, and the electron trigger efficiency is greater than 80\% and increases with momentum. 
The same trigger selection criteria are applied to simulated events, and scale factors ensure that the trigger 
efficiency in simulation matches that in data.

A further correction factor is applied to simulated events to account for a trigger inefficiency in 2016 and 2017, 
caused by a timing shift in the ECAL. An issue in some HCAL modules during 2018 resulted in a loss of efficiency 
in a specific $\eta$-$\phi$ region of the detector. Therefore, any event in data that has a muon, electron, or jet 
in the affected region during this period in time is rejected. The effect is accounted for in MC simulations using event weights.

To select events matching the expected topology of the signal process, we require events to have a single lepton \Pell
(muon or electron) within $\abs{\eta} < 2.4$, and $\pt > 30$ or 40\GeV for muons or electrons, respectively. 
Events are assigned to a muon or an electron channel based on the flavor of the selected lepton. 
Muons or electrons with \pt smaller than 55 or 120\GeV, respectively, are referred to as low-\pt leptons. 
Low-\pt muons need to fulfill $\Irel < 0.15$.
For low-\pt electrons, isolation is ensured by including relevant variables in the multivariate electron selection.
High-\pt leptons with \pt above these thresholds are required to satisfy one of two looser isolation criteria:
either $\DR (\Pell, \text{jet}) > 0.4$ or $p_{\text{T,rel}} > 25\GeV$, 
to suppress contributions from QCD multijet production~\cite{CMS:2018rkg}. 
The angular distance $\DR (\Pell, \text{jet})$ is calculated between the selected lepton and the closest small-radius 
jet in the $\eta$-$\phi$ plane, and $p_{\text{T,rel}}$ is the \pt component of the lepton perpendicular to the axis of the 
closest small-radius jet.

For a hadronically decaying top quark, we require at least one variable-radius jet with 
$\pt > 200\GeV$ and $\abs{\eta} < 2.4$, aiming to reconstruct the decay products in a single jet.
Additionally, events must contain at least four small-radius 
jets with $\pt > 30\GeV$ and $\abs{\eta} < 2.4$, at least one of which must be a \PQb jet.
In this selection, the variable-radius jets and small-radius jets are treated separately and 
are independent of each other. 
We do not test for a potential overlap between the small-radius and variable-radius jets in an event. 
Furthermore, $\ptmiss > 50\GeV$ is required to account for the neutrino from the \PW boson decay.

A scalar momentum sum variable \ST is defined as
\begin{equation}
    \ST = {\pt}^{\Pell} + \ptmiss +
    \sum_{
        \substack{ i }
    } {p_{\text{T},i}}^{\text{jet}},
\end{equation}
where the sum runs over all variable-radius jets with $\pt > 200\GeV$ and $\abs{\eta} < 2.4$. 
This variable is useful for discriminating signal from SM background events, as 
\tstartstar production results in a higher value of \ST on average. 
In Section \ref{sec:results}, we use the measured distribution in \ST for the statistical interpretation of the results. 
All events are required to have $\ST > 500\GeV$ to further reduce trigger inefficiency effects from events with low-\pt leptons and jets.

\begin{figure}[tbp]
    \centering
    \includegraphics[width=\cmsFigWidth]{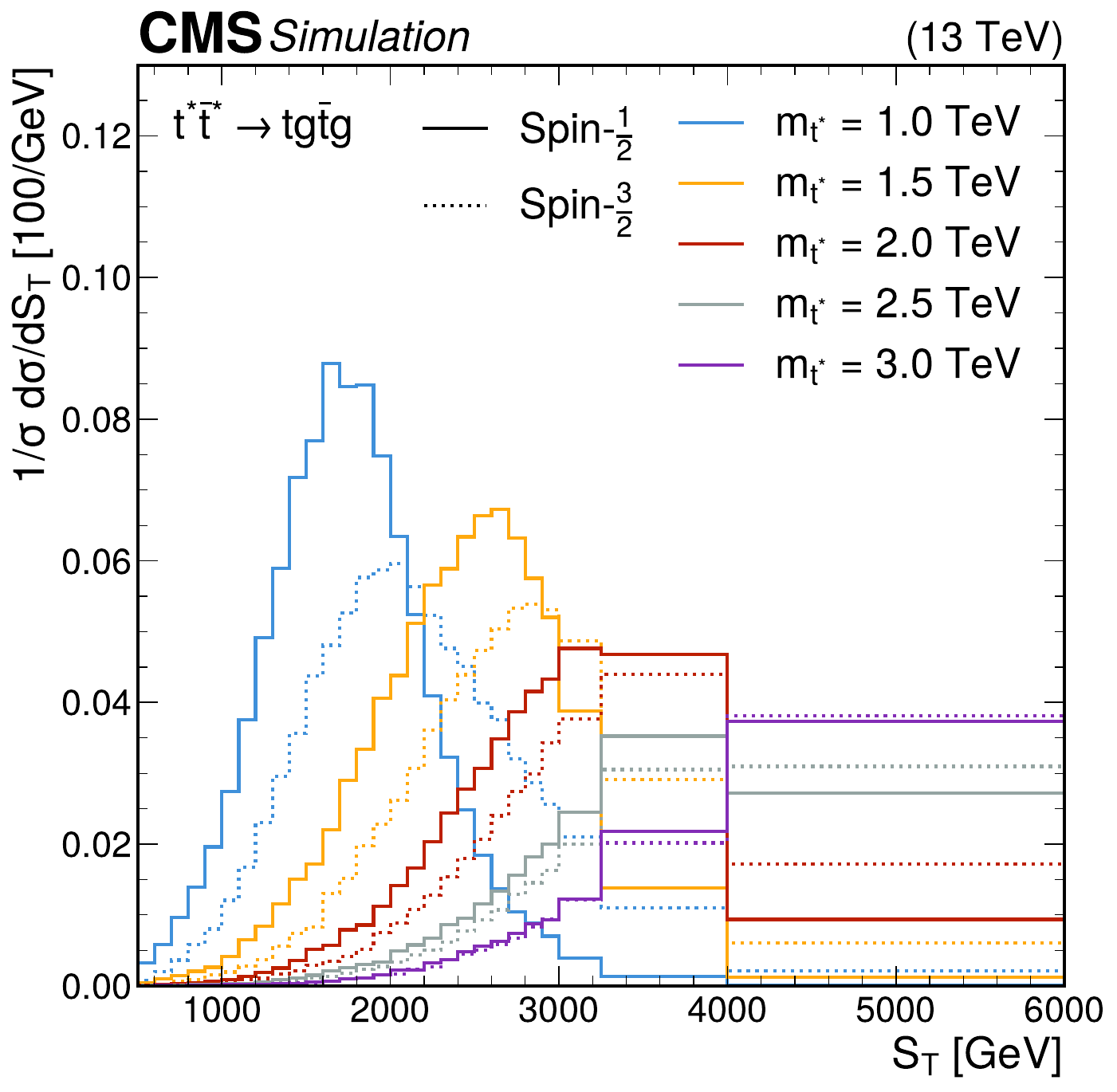}
    \caption{Distributions in \ST for \tstartstar signal samples with different simulated values of \mtstar, 
    for spin-1/2 (solid lines) and spin-3/2 (dashed lines) resonances.
    Each signal distribution is normalized to 100.
    \label{fig:signals_SR}}
\end{figure}

Depending on the spin and mass of the \tstar, around 50\% of signal events remain after the above requirements,
with higher efficiency for spin-3/2 compared to spin-1/2 \tstar, and increasing efficiency towards higher \mtstar.
In contrast, the contribution of SM background processes is substantially reduced, with efficiencies below 1\%. 
In Fig.~\ref{fig:signals_SR} we show the \ST distributions of various simulated signal samples. 
The spin-3/2 signals feature a peak in \ST around $2\mtstar$, shifted to smaller values for spin-1/2 signals. 
The spin-3/2 signals have higher \ST on average, a difference that is more pronounced at 
lower \tstar masses. This behavior results from higher \tstar momentum 
and invariant mass of the \tstartstar system for spin-3/2 signals compared to spin-1/2. 
Because the SM background contributions monotonically decrease with increasing \ST, 
we expect better sensitivity for spin-3/2 signals compared to spin-1/2. These differences become 
smaller for higher \tstar masses.

\section{Event classification}

{\tolerance=800
We use a DNN to classify events as signal like and background like. It is designed to output a single DNN score \sDNN, 
where signal-like events are given a high score and background-like events a low score. 
We pass 33 input features to the DNN, which are derived from the reconstructed PF candidates in signal events: 
\par}
\begin{itemize}
    \item \pt, $\eta$, $\phi$, and \Irel of the lepton, 
    \item \pt, $\eta$, $\phi$, the number of subjets, $\tau_1$, $\tau_2$, and $\tau_3$ of the three \pt-leading variable-radius jets,
    \item \pt, $\eta$, $\phi$ and the \textsc{DeepJet} score of the small-radius jet with the highest \textsc{DeepJet} score,
    \item \ptmiss and $\phi$ of \ptvecmiss, and 
    \item the multiplicities of the small-radius and variable-radius jets in the event. 
\end{itemize}
Zero padding is used in case features cannot be defined, for example for events containing less than three variable-radius jets.
We find that jet multiplicity and substructure variables are especially relevant for the DNN performance, 
together with information on the variable-radius jet that is third in the \pt hierarchy.
The DNN uses a fully connected feed-forward architecture with four hidden layers and 25 nodes each.
We trained it on simulated \ttbar events and a balanced mixture of spin-1/2 signal events with \mtstar between 700 and 3000\GeV. 
We verified that no \mtstar bias arises, by investigating the effect of repeating the training with one \mtstar sample removed.
The DNN is used for both spin scenarios, as we observe no substantial sensitivity increase when testing a separate, dedicated DNN for the spin-3/2 signals.

\begin{figure}[tb]
    \centering
    \includegraphics[width=\cmsFigWidth]{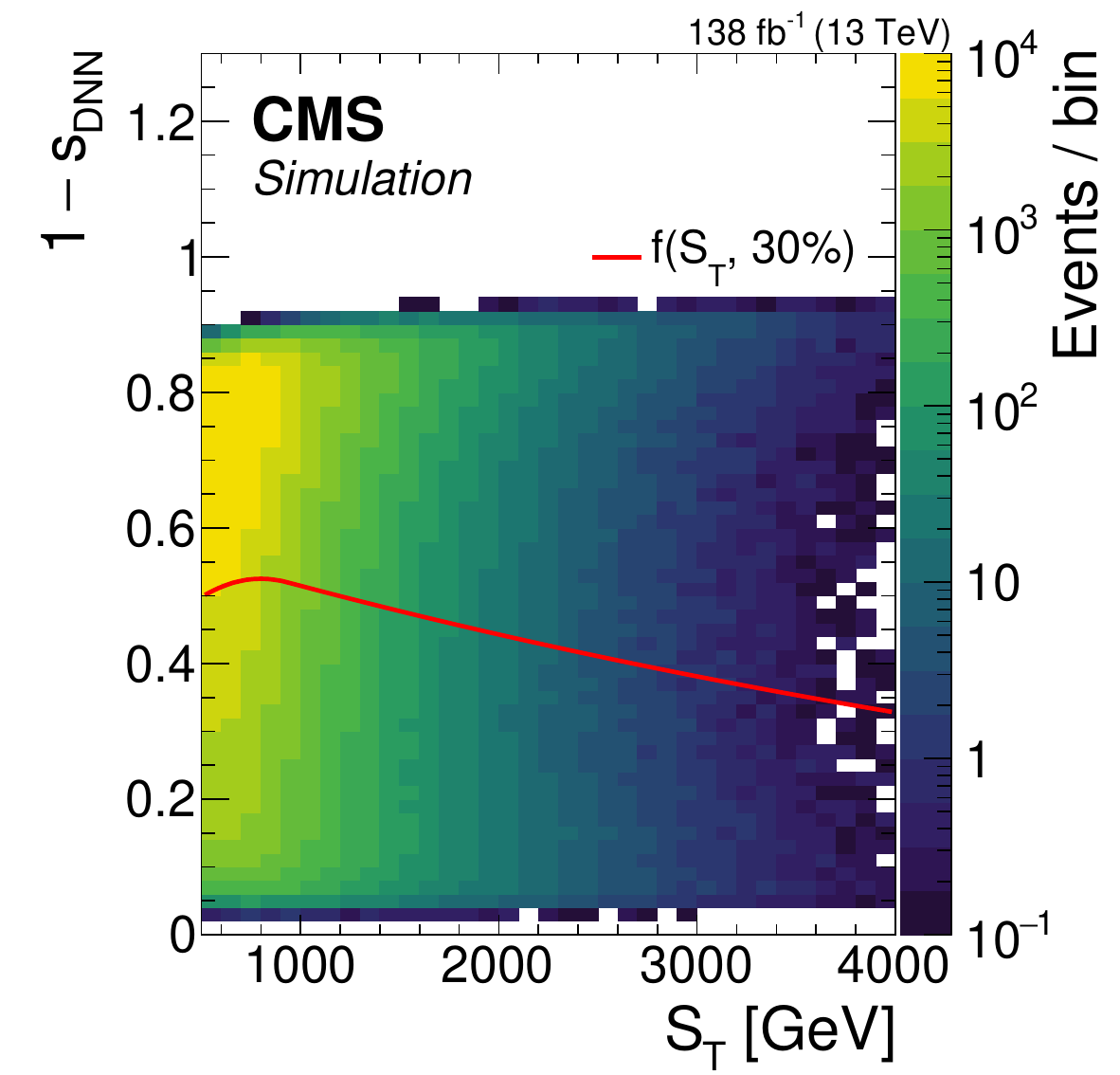}
    \caption{Two-dimensional distribution in $1 - \sDNN$ versus \ST for simulated \ttbar events. 
    The function $f(\ST, 30\%)$ (red line) is determined by specifying a 30\% selection efficiency for \ttbar events, 
    \ie, 30\% of the \ttbar events are below this function in each bin of \ST. 
    \label{fig:DDT}
    }
\end{figure}
Some of the DNN input variables are used to calculate \ST, which results in an output strongly correlated with \ST. 
To prevent this and coerce the network to learn auxiliary features that differ between signal and background events, 
background events are reweighted to have an \ST distribution identical to that of the signal events used in the training. 
Aside from ensuring balance of the signal and background classes during training,
this reduces the correlation between \ST and \sDNN.
At the same time, it facilitates the exploitation of information other than the \pt, 
such as angular correlations, jet multiplicity, and the substructure of variable-radius jets. 
The importance of these features for the network decision increases after the \ST reweighting.
However, some correlation between \ST and \sDNN remains, which is alleviated in a second decorrelation step. 
We use the ``designing decorrelated taggers'' (DDT) technique~\cite{Dolen:2016kst}, where we shift \sDNN 
by an \ST-dependent value $f(\ST, \eptt)$ for each event, 
\begin{equation}
    \sDDT = \sDNN - f(\ST, \eptt).
\end{equation}
To determine the function $f(\ST, \eptt)$, a set of \sDNN selection thresholds are found in intervals of 100\GeV in \ST. 
In each interval, the threshold value corresponds to a given efficiency \eptt for selecting \ttbar background events.
To obtain the continuous function $f(\ST, \eptt)$, we fit the selection thresholds using an analytic function.
The two-dimensional distribution in \ST versus $1-\sDNN$ and the resulting function 
$f(\ST, \eptt)$ are shown in Fig. \ref{fig:DDT} for $\eptt = 30\%$.

The resulting score \sDDT is uncorrelated with \ST, \ie, a selection based on \sDDT results in a 
constant \ttbar background efficiency as a function of \ST. This is crucial for the background estimation described below.  
We define a signal region (SR) with $\sDDT>0$ and a validation region (VR) with $\sDDT<0$. 
While the SR is used in the statistical analysis in the search for a signal, the VR is only used to validate the 
background estimation presented below. We find that a \ttbar selection efficiency $\eptt = 30\%$ results in the best 
sensitivity of this analysis, while ensuring that the split of events between the SR and VR 
only minimally changes the shape of the \ST distributions for the SM backgrounds. 
By construction, the SR contains 30\% of all \ttbar events after the analysis selection and the VR contains 70\%. 
The efficiency for signal events to enter the SR varies between 55 and 75\%, depending on the 
spin and \mtstar.

\section{Background estimation}

The SM processes that can result in a final state similar to the one expected from signal events 
can be divided into two classes: backgrounds containing top quarks, \ttbar and single top quark events, 
and backgrounds without on-shell top quarks (non-\PQt backgrounds), mostly consisting of events from 
\PW{}+jets, Drell--Yan, QCD multijet, and diboson production.

We estimate backgrounds with top quarks using simulation, for which the correction factors described in Section~\ref{sec:eventreconstruction}
ensure a good description of the data. In addition to these corrections, a reweighting as a function 
of the parton-level \pt of the top quark is applied to background events. These weights correct for a top quark 
\pt-dependent mismodeling in simulations at NLO precision~\cite{toppt1, toppt2}. 

\begin{figure}[tbp]
    \centering
    \includegraphics[width=\cmsFigWidth]{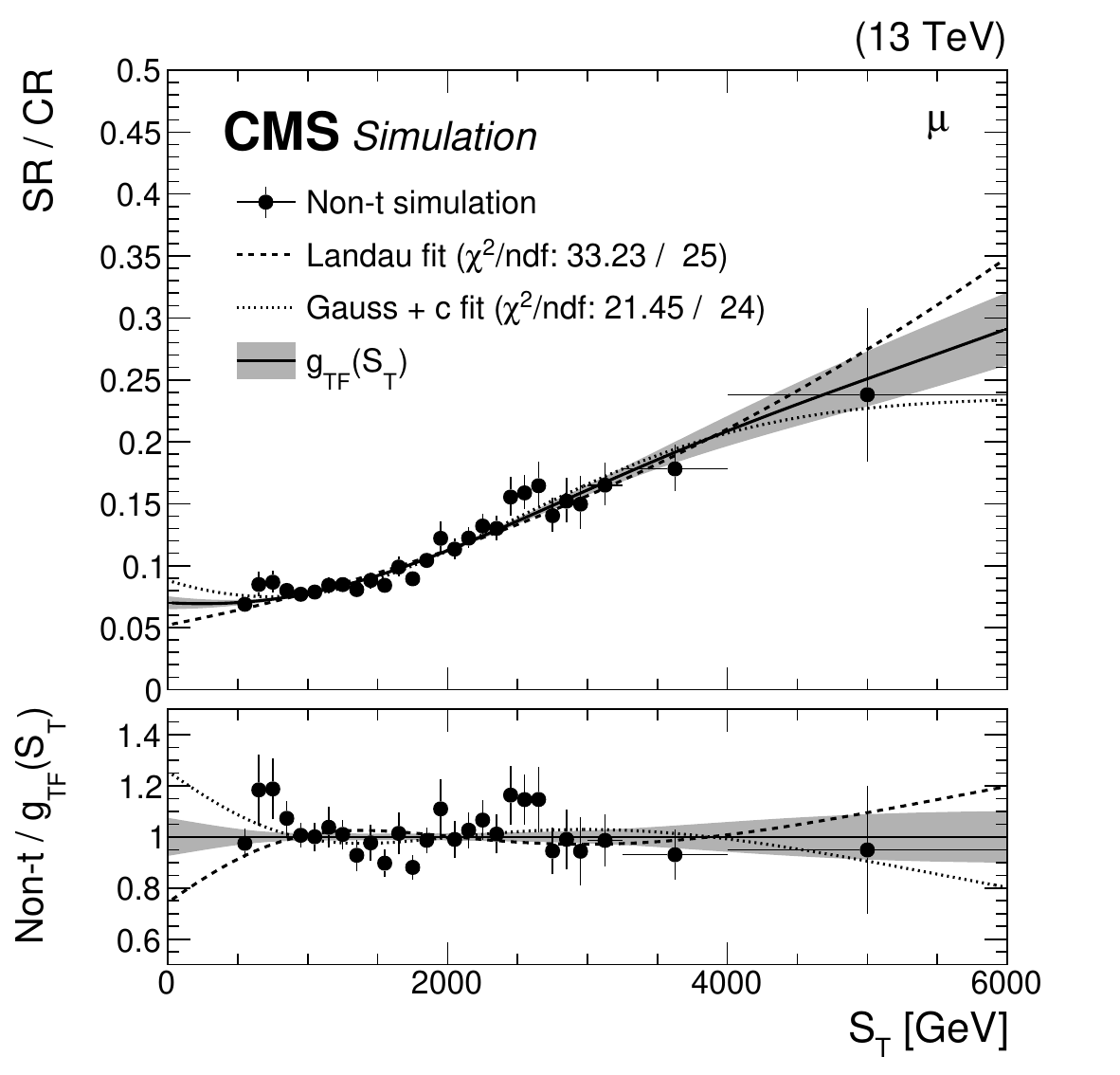}
    \includegraphics[width=\cmsFigWidth]{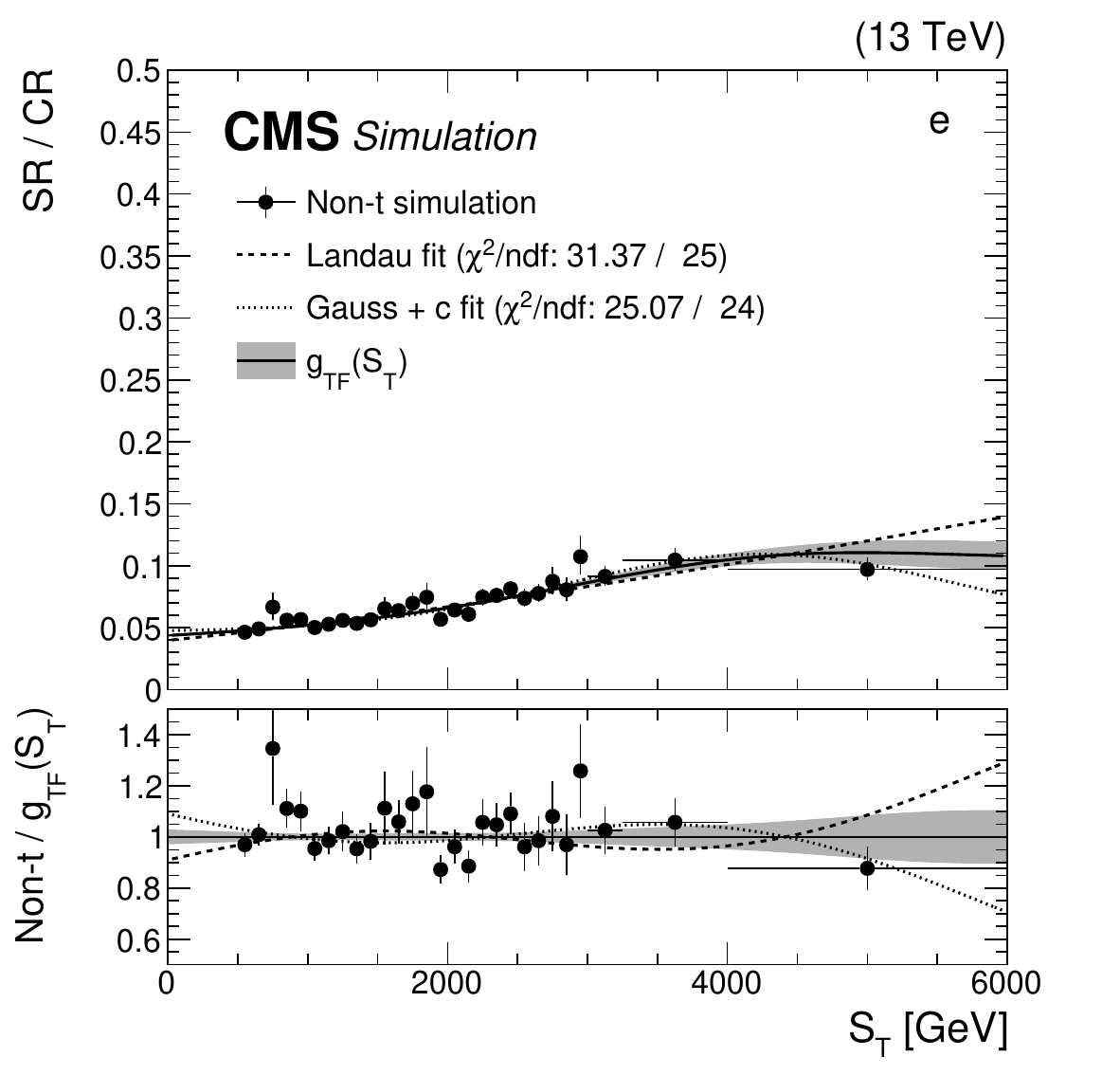}
    \caption{The simulation-based ratios between the \ST distributions in the SRs and CRs for the muon (\cmsLeft) 
    and electron (\cmsRight) channels. 
    Two functions are fit to each ratio, and the final transfer function used for the non-\PQt 
    background estimation is taken to be their average. 
    The statistical uncertainties in the transfer functions are shown as grey bands.
    In the lower panels, the simulation-based ratios, fit functions, and statistical uncertainties are shown relative to the final transfer function. 
    \label{fig:alpharatio}
    }
\end{figure}

{\tolerance=1200
We estimate the contribution from backgrounds without top quarks with a method based on control samples in data, 
because the production of weak vector bosons with multiple jets and QCD multijet 
production are not modeled with sufficient accuracy. 
We define a control region (CR) that contains events passing all the requirements described in 
Section~\ref{sec:selection}, except for the \PQb jet selection. Only events without a \PQb jet enter the CR. 
No requirement on \sDDT is imposed in the CR. The CR is enriched in non-\PQt backgrounds,
with top quark background contaminations amounting to 10--20\%, and is statistically 
independent of the SR. We use simulated events to construct an \ST-dependent ratio between the non-\PQt background 
contributions in the CR and SR. We fit the ratio with two functions, a Landau distribution and a 
Gaussian distribution with a constant offset. The average of these two functions then defines a 
transfer function $\gTF(\ST)$. The ratios between the \ST distributions in the SR and CR 
are shown in Fig.~\ref{fig:alpharatio}, together with the fits of the transfer functions. 
\par}

\begin{figure}[tbp]
    \centering
    \includegraphics[width=\cmsFigWidth]{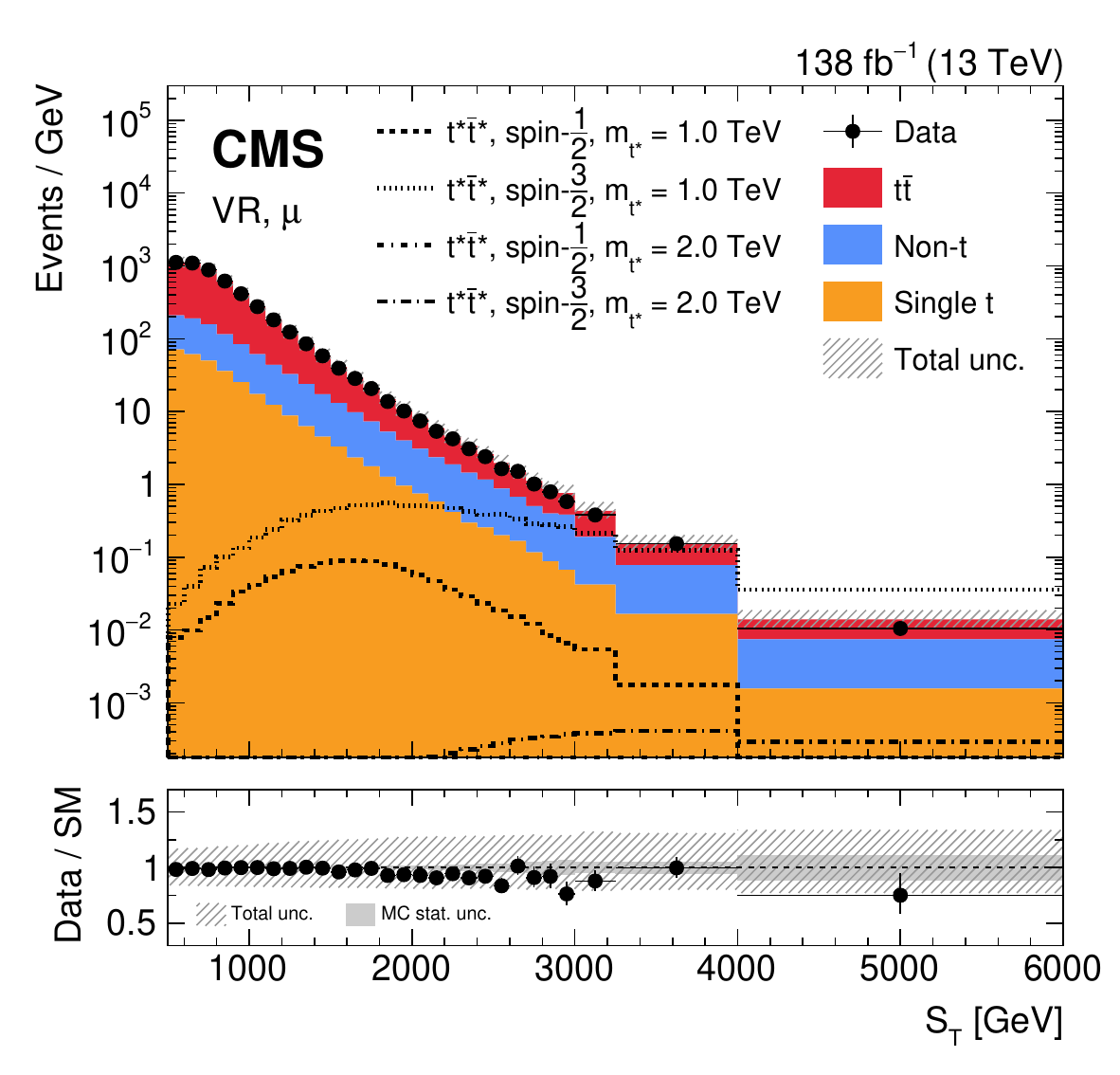}
    \includegraphics[width=\cmsFigWidth]{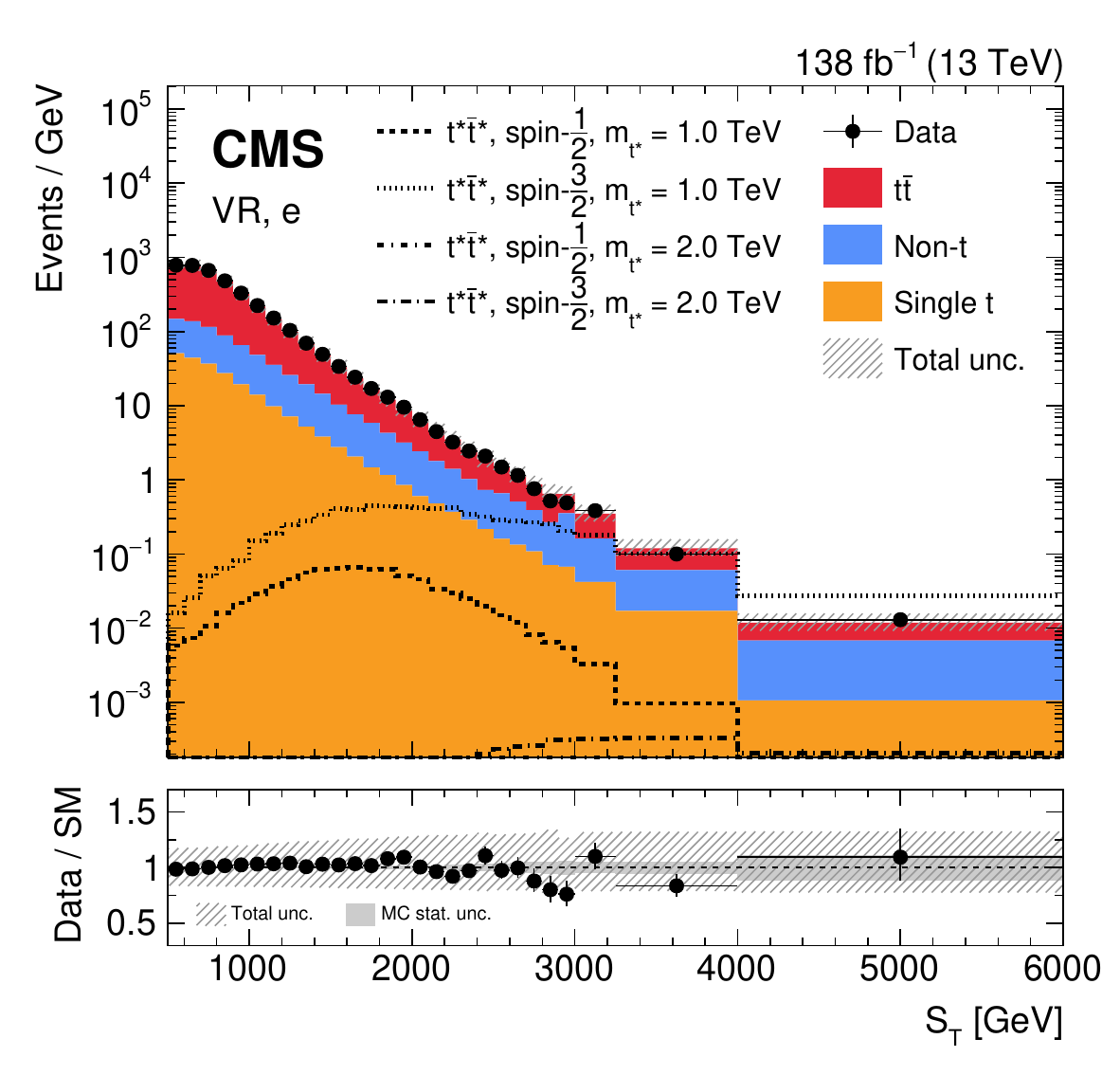}
    \caption{Distributions of \ST in the VR for the muon (\cmsLeft) and electron (\cmsRight) channels. 
    The total uncertainties are shown as hatched bands.
    The signal distributions are scaled to the cross sections predicted by theory. 
    Ratios of data to the expected backgrounds are shown in the lower panels. 
    \label{fig:validationregion}
    }
\end{figure}

The function $\gTF(\ST)$ is used to estimate the non-\PQt background contribution in the SR as
\ifthenelse{\boolean{cms@external}}{
\begin{multline}
    N_{\text{SR}}( \text{non-\PQt bkg} ) = \\ \gTF(\ST) \left( N_{ \text{CR} }(\text{data}) - N_{ \text{CR} }(\text{top bkg}) \right), 
\end{multline}
}{
\begin{equation}
    N_{\text{SR}}( \text{non-\PQt bkg} ) = \gTF(\ST) \left( N_{ \text{CR} }(\text{data}) - N_{ \text{CR} }(\text{top bkg}) \right), 
\end{equation}
} 
{\tolerance=800
where $N_{ \text{CR} }(\text{data})$ and $N_{ \text{CR} }(\text{top bkg})$ are the event counts 
in the CR for data and for the simulated samples with on-shell top quarks, respectively. 
The event counts are obtained in intervals of \ST. 
For the muon and electron channels, the number of events in the CR is multiplied by $\gTF(\ST)$,
after subtracting contributions of simulated top quark backgrounds.
Since the composition of non-\PQt backgrounds differs between the muon and electron channels,
the method is applied to each channel individually.
\par}
{\tolerance=800
We verify the estimation of the background in the SR using the VR.
Results from the background estimation for the VR are shown in Fig.~\ref{fig:validationregion} 
for the muon and electron channels.
The bin widths increase with increasing \ST so that each bin is appropriately populated in data and simulation.
The uncertainty arising from the limited size of the simulated background samples becomes non-negligible towards high \ST,
but its impact on the sensitivity of the analysis is small compared to other uncertainties.
The predicted backgrounds describe the measured data within the uncertainties in both channels.
\par}

\section{Systematic uncertainties}
\label{sec:uncertainties}

Several systematic effects introduce uncertainties affecting the normalization and shape of distributions in \ST. 
We summarize these in the following, split into experimental and theoretical uncertainties. 
All uncertainties, which can affect either the rate or both the rate and the shape of \ST,
are included as nuisance parameters in the statistical analysis presented below.
Details on the implementation of nuisance parameters are provided in Ref.~\cite{combine}.
To give an estimate of the relevance of the systematic uncertainties, we provide an approximate size of each source 
for the largest SM contribution it affects.

\subsection{Experimental uncertainties}

Systematic uncertainties arise from the selection of \PQb jets and the corresponding efficiencies. 
The \PQb jet selection uncertainties are split into different sources, depending on the jets they affect: 
jets originating from \PQb quarks, \PQc quarks, or light jets.
Furthermore, these are split into statistical and systematic components, 
where the former are not correlated between eras of data taking, but the latter are. The uncertainty from each source corresponds to variations 
of up to 10\% for \ttbar events. An additional \PQb tagging related uncertainty is connected to normalization 
differences before and after applying the \PQb tagging correction factors. 
Separate corrections to the normalization are derived for the electron and muon channels. 
The differences between these and the correction obtained for a combination of the two channels 
is taken to be the associated uncertainty.
This uncertainty can reach up to 5\% for simulated \ttbar events 
and is correlated between the years of data taking.

The impact of uncertainties associated with the jet energy scale and jet energy resolution corrections 
on small radius jets, variable-radius jets and \ptvecmiss is estimated by varying the 
corresponding corrections within their uncertainties.
The resulting uncertainty is correlated between all affected objects and treated as uncorrelated between years.
This uncertainty has an impact of up to 4\% in \ttbar events. 

Uncertainties connected to correction factors for the lepton reconstruction, identification, isolation, and triggers 
are of the order of 1\% or smaller for \ttbar events, and are not correlated between years of data taking.
An uncertainty in the correction factor accounting for the 2016 and 2017 trigger inefficiency is considered as well.

The total inelastic cross section of 69.2\unit{mb}, used to correct the pileup distribution in simulation, 
is varied by ${\pm}4.6\%$~\cite{CMS:2018mlc}, resulting in an uncertainty of around 2\% for \ttbar events.

A total uncertainty of 1.6\% in the integrated luminosity of 138\fbinv is considered~\cite{CMS-LUM-17-003, CMS-PAS-LUM-17-004, CMS-PAS-LUM-18-002}. 
This uncertainty affects only the normalization of simulated samples. 

Another source of uncertainty, specific to this analysis, arises from the DNN decorrelation procedure. 
We take the residual differences between the shapes of the \ST distributions in the SR and VR for top quark backgrounds 
as an estimate of the associated uncertainty. The resultant impact corresponds to variations of up to 10\% for \ttbar events. 

Separate uncertainties are considered on the results of the data-driven background estimation.
The procedure is repeated for each of the two functions used to construct $\gTF(\ST)$. 
The resulting distributions are used as a systematic variation around the nominal non-\PQt background prediction. 
The effect is treated as fully correlated between years and is below 8\% for the non-\PQt background. 
The effect of the statistical uncertainty in $\gTF(\ST)$ amounts to around 5\%. 
We estimate the systematic uncertainty connected to the \PQb tagging veto in the definition of the CR 
by repeating the background estimation for variations of the \PQb tagging corrections, where the 
individual sources are combined. The resulting uncertainty can reach up to 15\% for the non-\PQt background 
and is treated as fully correlated between years.

In summary, the dominant experimental systematic uncertainties in this analysis are from the \PQb tagging corrections 
and the decorrelation procedure for the background estimation.

\subsection{Theoretical uncertainties}

The largest systematic uncertainty in the modeling of SM background processes results from missing higher-order effects
in the simulation. We estimate this effect by varying the renormalization and factorization scales 
in the simulation by factors of 2 and 1/2. An envelope is constructed from the possible variations, 
except for variations where one scale is varied by 2 and the other by 1/2. 
This uncertainty is taken to be fully correlated between data-taking years. It includes a normalization effect 
covering uncertainties in the production cross sections of the SM background processes and varies 
between 20 and 60\% for \ttbar events, with the larger values at high \ST. For signal samples, 
we remove uncertainties in the normalization from scale variations and consider only acceptance and shape effects. 

The uncertainty in the correction of the \pt of the parton-level top quark is estimated
from the difference in the \ST distribution before and after applying this correction.
For \ttbar events, it amounts to below 20\%.

An uncertainty resulting from the choice of PDFs is obtained using the standard deviation of 100 different MC replicas, 
following the procedure described in Ref.~\cite{PDFunc}. 
The uncertainty has an unimportant effect on the analysis.
For signal samples, only the acceptance and shape effects are considered for the PDF uncertainties.

Other sources of uncertainties related to the modelling of background processes,
like uncertainties in the simulation of initial and final state radiation,
the matching of matrix elements to the parton shower,
or the modelling of hadronization,
are small compared to the ones considered here.

\section{Results}
\label{sec:results}

The \ST distribution in the SR, measured in the muon and electron channels, is the main observable of this analysis. 
The backgrounds from \ttbar and single top quark 
production are taken from simulation, with the appropriate correction factors applied. The \ST distributions of 
the non-\PQt backgrounds are extrapolated from the CR. 
Signal distributions are taken from simulations. We combine all years of data and include 
the uncertainties described in Section~\ref{sec:uncertainties} as nuisance parameters.

A statistical analysis is performed to probe for the existence of \tstartstar production 
with a binned maximum likelihood fit. 
The modified frequentist approach~\cite{CMS-NOTE-2011-005,CLS1,CLS2},
known as the \CLs criterion with the profile likelihood ratio as the test statistic, 
is used in this search to set 95\% confidence level (\CL) upper limits on the product of the 
production cross section for $\pp\to\tstartstar$ and the branching fraction squared $\BR^2(\tstartg)$.
We use the asymptotic approximation to the profile likelihood test statistic~\cite{Cowan:2010js}.
The statistical analysis is performed using the CMS statistical analysis tool \textsc{Combine}~\cite{combine}, 
which is based on the \textsc{RooFit}~\cite{Verkerke:2003ir} and \textsc{RooStats}~\cite{Moneta:2010pm} frameworks.

\begin{figure}[tbp]
    \centering
    \includegraphics[width=\cmsFigWidth]{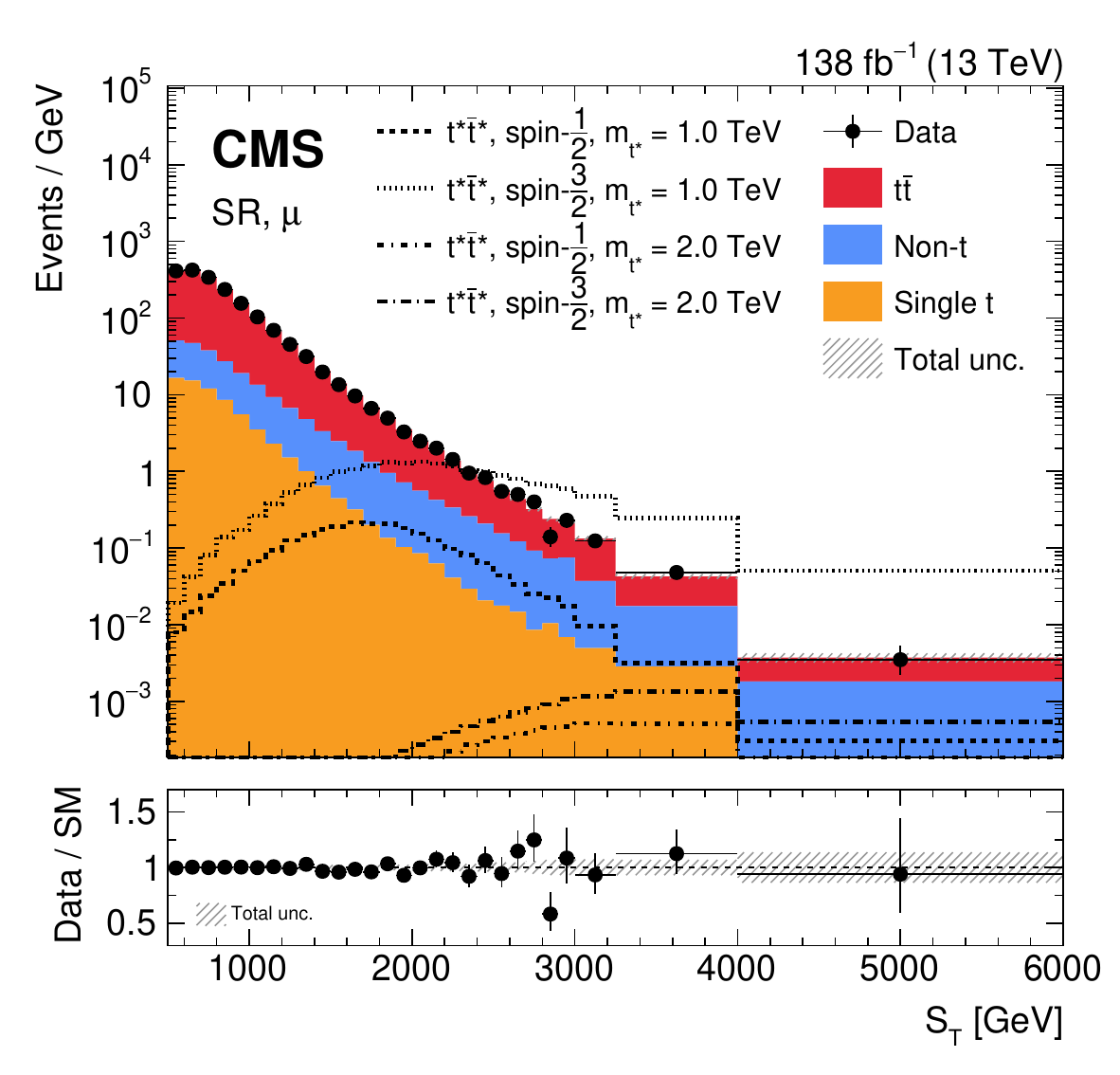}
    \includegraphics[width=\cmsFigWidth]{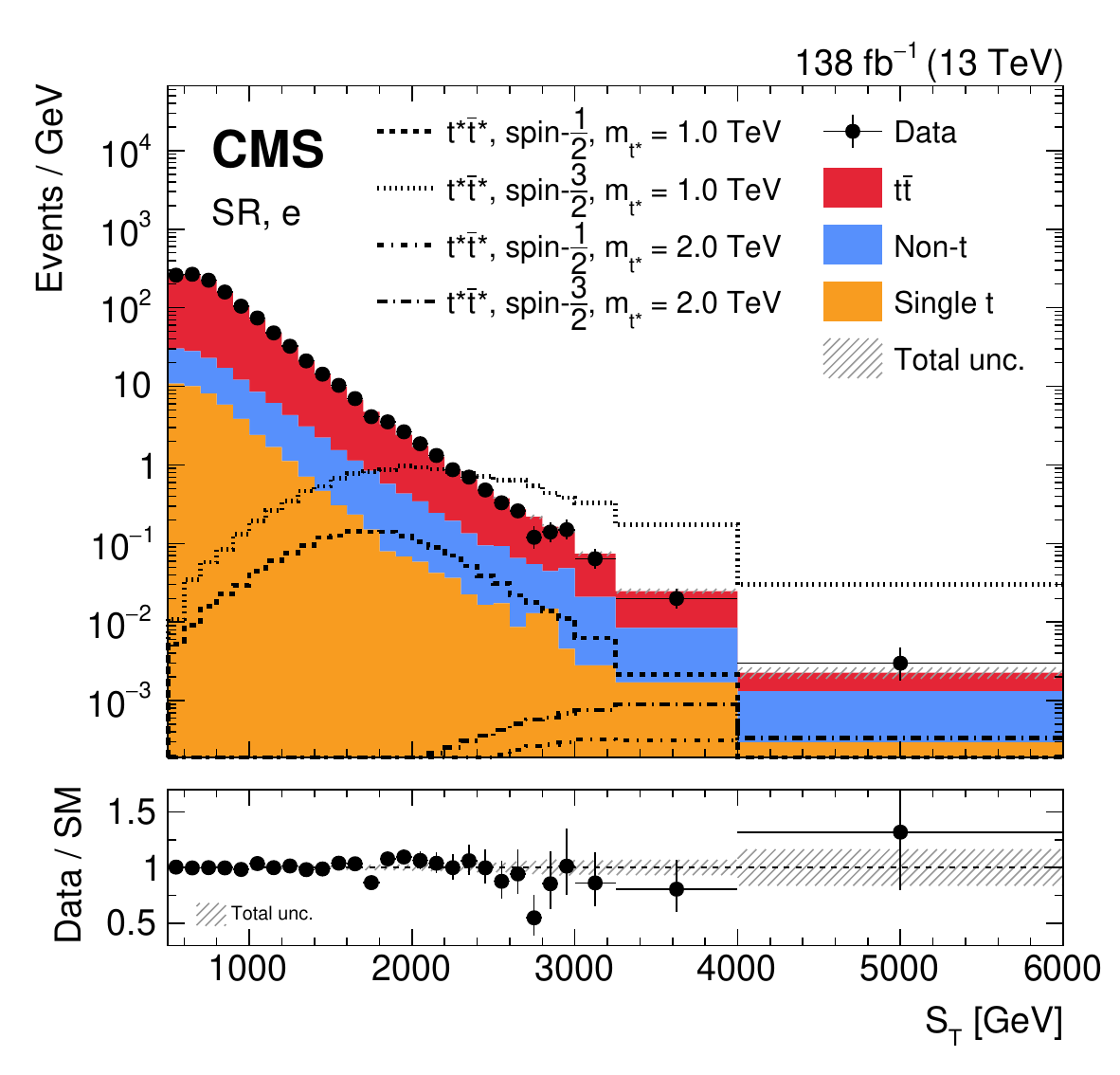}
    \caption{Distributions in \ST in the SR for the muon (\cmsLeft) and electron (\cmsRight) channels, after a background-only fit to the data. 
    The signal distributions are scaled to the cross section predicted by the theory. 
    The hatched bands show the post-fit uncertainty band, combining all sources of uncertainty. 
    The ratio of data to the background predictions is shown in the panels below the distributions. 
    \label{fig:signalregion}
    }
\end{figure}
\begin{figure*}[tbp]
    \centering
    \includegraphics[width=0.8\textwidth]{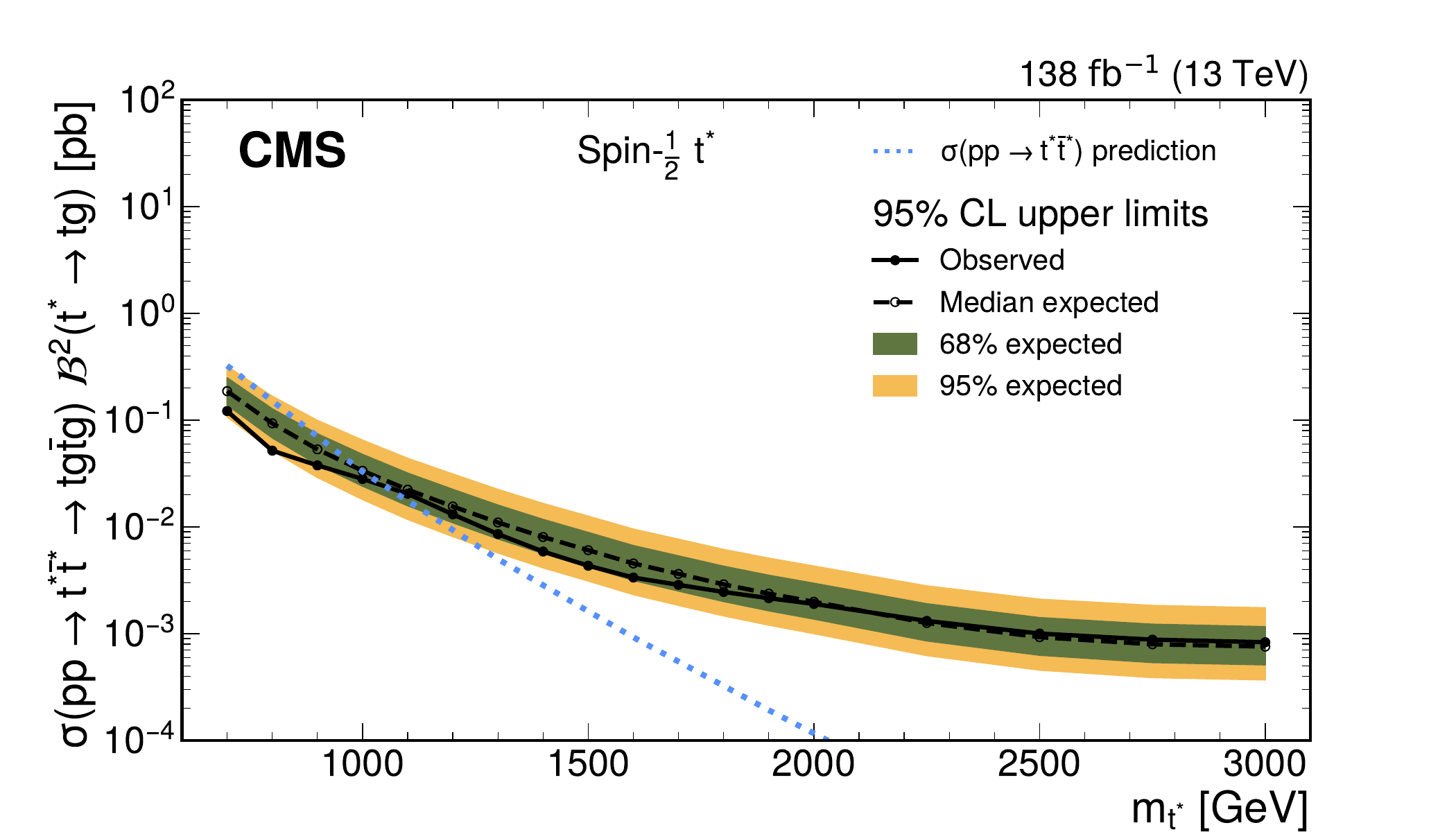}
    \caption{Expected and observed 95\% \CL upper limits on the product of the \tstartstar production cross section  
    and the branching fraction squared $\BR^2(\tstartg)$ for a spin-1/2 \tstar as a function of \mtstar. 
    The inner (green) and outer (yellow) bands give the central probability intervals containing 
    68 and 95\% of the expected upper limits under the background-only hypothesis. 
    The cross section predicted by theory, following the EFT approach introduced in Ref.~\cite{Alhazmi:2018whk}, 
    is shown as a dotted line, assuming $\BR(\tstartg)=1$.
    \label{fig:limits12}
    }
\end{figure*}
\begin{figure*}[tbp]
    \centering
    \includegraphics[width=0.8\textwidth]{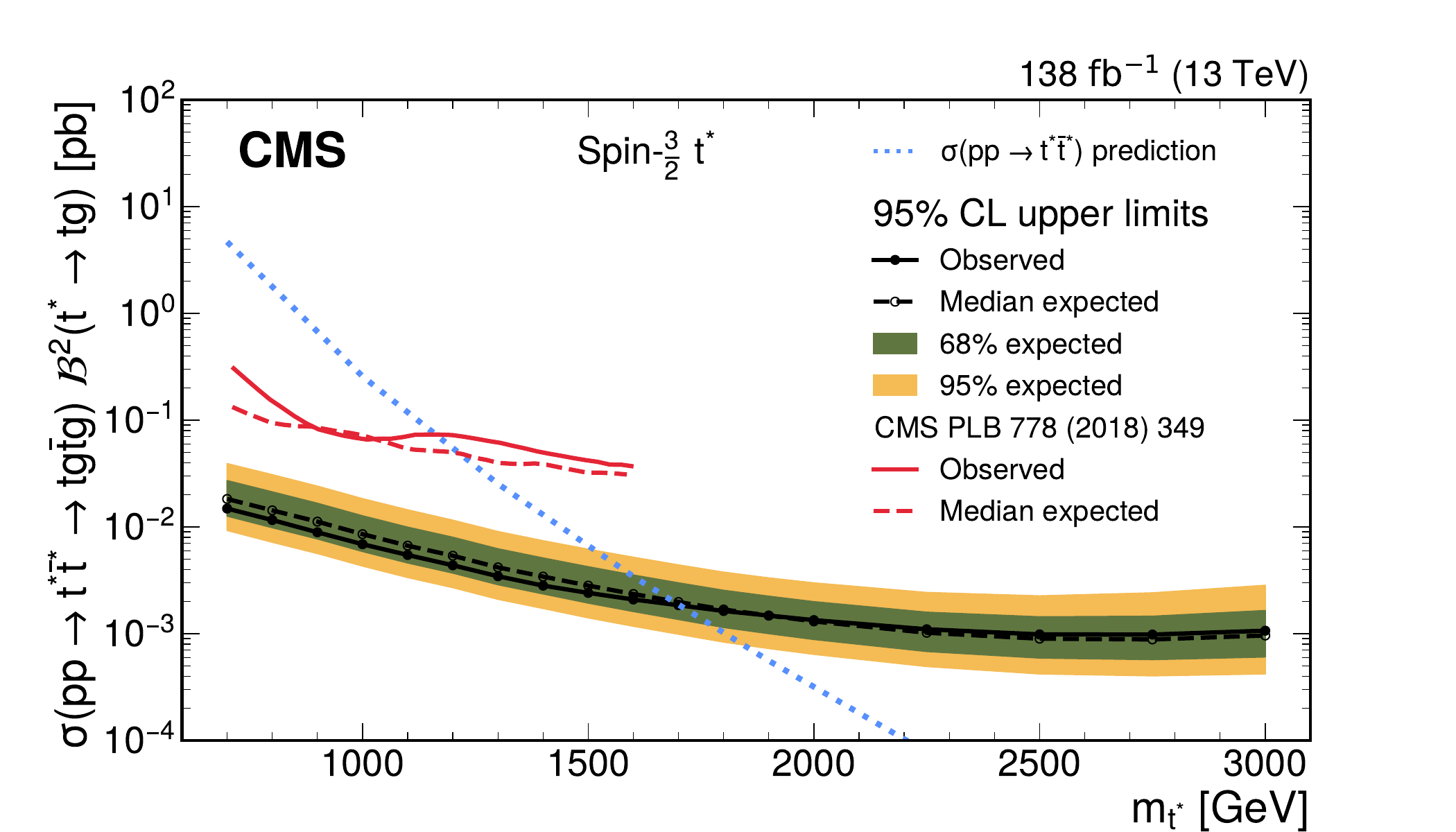}
    \caption{Expected and observed 95\% \CL upper limits on the product of the \tstartstar production cross section  
    and the branching fraction squared $\BR^2(\tstartg)$ for a spin-3/2 \tstar as a function of \mtstar. 
    The inner (green) and outer (yellow) bands give the central probability intervals containing 
    68 and 95\% of the expected upper limits under the background-only hypothesis. 
    The cross section predicted by theory, following the EFT approach introduced in Ref.~\cite{Alhazmi:2018whk}, 
    is shown as a dotted line, assuming $\BR(\tstartg)=1$.
    The results of the previous CMS analysis~\cite{CMS:2017ixp}, using data corresponding to an integrated luminosity of 
    35.9\fbinv, are shown as well.
    \label{fig:limits32}
    }
\end{figure*}

{\tolerance=1600
We show the \ST distributions in the SR for the muon and electron channels in Fig.~\ref{fig:signalregion} 
after the background-only maximum likelihood fit to the data. 
No significant deviation from the SM predictions is observed. The resulting upper limits on the product of the
production cross section and the branching fraction squared are presented in Figs.~\ref{fig:limits12} and \ref{fig:limits32} 
for spin-1/2 and spin-3/2 resonances, respectively. For the spin-1/2 case, the limits are found to be 
between 120\unit{fb} (190\unit{fb} expected) for $\mtstar=700\GeV$, and 0.8\unit{fb} (0.8\unit{fb} expected) 
for $\mtstar=3000\GeV$. For the spin-3/2 case, the upper limits are between 15\unit{fb} (18\unit{fb} expected) 
at 700\GeV and 1.0\unit{fb} (0.9\unit{fb} expected) at 2750\GeV.
\par}
As expected from the differences in the \ST distributions between spin-1/2 and spin-3/2 \tstar signals,
the analysis is more sensitive to spin-3/2 signals at low \mtstar.
At high \mtstar, the sensitivity is comparable, with the analysis being slightly more sensitive to spin-1/2 signals.
This is a result of the efficiency of the DNN-based SR definition, which decreases with increasing \mtstar.
This effect is more pronounced for spin-3/2 than for spin-1/2 \tstar signals.

For both spin scenarios, we compare the results to theory predictions of the \tstartstar pair production cross section. 
These are determined as described in Section~\ref{sec:samples}, using an EFT approach to describe the interaction of the 
\tstar and SM particles. We assume a branching fraction of 100\% for the \tstartg decay.

The existence of a spin-1/2 \tstar is excluded up to \mtstar values of 1050\GeV, with 990\GeV expected. 
These are the first mass exclusion limits on a spin-1/2 \tstar in the \tstartg decay channel at 13 \TeV, 
made possible by the improvements in this analysis compared to the previous analysis~\cite{CMS:2017ixp}. 
For the spin-3/2 case, the lower mass limit is considerably higher because of the higher predicted cross 
section and the better sensitivity of the analysis to spin-3/2 \tstar resonances below 2\TeV.
The existence of a spin-3/2 \tstar is excluded below \mtstar of 1700\GeV, with 1690\GeV expected.

When comparing the exclusion limits to the ones from the previous CMS result, based on 13\TeV data 
corresponding to an integrated luminosity of 35.9\fbinv, a substantial improvement is visible. 
The increased size of the analyzed data set yields an improvement in sensitivity to the cross section limit of about a factor of two. 
In addition, the sensitivity is increased by improvements of the analysis strategy. 
We impose looser lepton and jet multiplicity requirements and use variable-radius jets to access the 
Lorentz-boosted \tstar decay products. This, combined with the addition of an event classification DNN 
and the usage of \ST instead of a reconstruction of the \tstar system, leads to an increased signal efficiency: 
the fraction of signal events reaching the SR is about five times higher compared to the previous analysis. 
In contrast, background yields in the SR only increase by about 10\%. 
Overall, we achieve an increase in sensitivity by about a factor of five on top of the factor of two from the larger 
data set, resulting in a total improvement in sensitivity by about an order of magnitude compared to the previous analysis.

\section{Summary}

A search for the pair production of heavy top-quark partners \tstar has been presented, where the \tstar couples 
predominantly to gluons and decays to a top quark and a gluon, \tstartg. Both spin-1/2 and spin-3/2 resonances 
are considered. The analysis uses 13\TeV proton-proton collision data collected by the CMS experiment 
between 2016 and 2018, corresponding to an integrated luminosity of 138\fbinv. 
The final state analyzed consists of a lepton with high transverse momentum, missing transverse momentum 
and several jets. 
A deep neural network is used to identify potential signal events. 
With a two-step decorrelation procedure, independence of the deep neural network output from the main observable \ST has been achieved,
where \ST is the scalar sum of the transverse momenta of the selected lepton and jets, 
and the missing transverse momentum. 
No statistically significant deviation from the background prediction was found. 
Upper limits at 95\% confidence level are derived on the product of the \tstartstar production cross section 
and branching fraction squared for \tstartg. 
These are between 120 and 0.8\unit{fb} for a spin-1/2 \tstar and between 
15 and 1.0\unit{fb} for a spin-3/2 \tstar, depending on the \tstar mass. 
A comparison of these limits with the theory predictions results in mass limits for the \tstar resonances, 
where the existence of a spin-1/2 \tstar is excluded below a mass of 1050\GeV and 
for a spin-3/2 \tstar below a mass of 1700\GeV. 
These are the most stringent limits to date and the first exclusion limit for a spin-1/2 \tstar resonance at 13 \TeV. 
The results also substantially improve the spin-3/2 exclusion limits compared to previous results.

\begin{acknowledgments}
We congratulate our colleagues in the CERN accelerator departments for the excellent performance of the LHC and thank the technical and administrative staffs at CERN and at other CMS institutes for their contributions to the success of the CMS effort. In addition, we gratefully acknowledge the computing cemters and personnel of the Worldwide LHC Computing Grid and other cemters for delivering so effectively the computing infrastructure essential to our analyses. Finally, we acknowledge the enduring support for the construction and operation of the LHC, the CMS detector, and the supporting computing infrastructure provided by the following funding agencies: SC (Armenia), BMBWF and FWF (Austria); FNRS and FWO (Belgium); CNPq, CAPES, FAPERJ, FAPERGS, and FAPESP (Brazil); MES and BNSF (Bulgaria); CERN; CAS, MoST, and NSFC (China); MINCIENCIAS (Colombia); MSES and CSF (Croatia); RIF (Cyprus); SENESCYT (Ecuador); ERC PRG, RVTT3 and MoER TK202 (Estonia); Academy of Finland, MEC, and HIP (Finland); CEA and CNRS/IN2P3 (France); SRNSF (Georgia); BMBF, DFG, and HGF (Germany); GSRI (Greece); NKFIH (Hungary); DAE and DST (India); IPM (Iran); SFI (Ireland); INFN (Italy); MSIP and NRF (Republic of Korea); MES (Latvia); LMTLT (Lithuania); MOE and UM (Malaysia); BUAP, CINVESTAV, CONACYT, LNS, SEP, and UASLP-FAI (Mexico); MOS (Montenegro); MBIE (New Zealand); PAEC (Pakistan); MES and NSC (Poland); FCT (Portugal);  MESTD (Serbia); MCIN/AEI and PCTI (Spain); MOSTR (Sri Lanka); Swiss Funding Agencies (Switzerland); MST (Taipei); MHESI and NSTDA (Thailand); TUBITAK and TENMAK (Turkey); NASU (Ukraine); STFC (United Kingdom); DOE and NSF (USA).

\hyphenation{Rachada-pisek} Individuals have received support from the Marie-Curie program and the European Research Council and Horizon 2020 Grant, contract Nos.\ 675440, 724704, 752730, 758316, 765710, 824093, 101115353, 101002207, and COST Action CA16108 (European Union); the Leventis Foundation; the Alfred P.\ Sloan Foundation; the Alexander von Humboldt Foundation; the Science Committee, project no. 22rl-037 (Armenia); the Belgian Federal Science Policy Office; the Fonds pour la Formation \`a la Recherche dans l'Industrie et dans l'Agriculture (FRIA-Belgium); the F.R.S.-FNRS and FWO (Belgium) under the ``Excellence of Science -- EOS" -- be.h project n.\ 30820817; the Beijing Municipal Science \& Technology Commission, No. Z191100007219010 and Fundamental Research Funds for the Central Universities (China); the Ministry of Education, Youth and Sports (MEYS) of the Czech Republic; the Shota Rustaveli National Science Foundation, grant FR-22-985 (Georgia); the Deutsche Forschungsgemeinschaft (DFG), among others, under Germany's Excellence Strategy -- EXC 2121 ``Quantum Universe" -- 390833306, and under project number 400140256 - GRK2497; the Hellenic Foundation for Research and Innovation (HFRI), Project Number 2288 (Greece); the Hungarian Academy of Sciences, the New National Excellence Program - \'UNKP, the NKFIH research grants K 131991, K 133046, K 138136, K 143460, K 143477, K 146913, K 146914, K 147048, 2020-2.2.1-ED-2021-00181, and TKP2021-NKTA-64 (Hungary); the Council of Science and Industrial Research, India; ICSC -- National Research Cemter for High Performance Computing, Big Data and Quantum Computing and FAIR -- Future Artificial Intelligence Research, funded by the NextGenerationEU program (Italy); the Latvian Council of Science; the Ministry of Education and Science, project no. 2022/WK/14, and the National Science Center, contracts Opus 2021/41/B/ST2/01369 and 2021/43/B/ST2/01552 (Poland); the Funda\c{c}\~ao para a Ci\^encia e a Tecnologia, grant CEECIND/01334/2018 (Portugal); the National Priorities Research Program by Qatar National Research Fund; MCIN/AEI/10.13039/501100011033, ERDF ``a way of making Europe", and the Programa Estatal de Fomento de la Investigaci{\'o}n Cient{\'i}fica y T{\'e}cnica de Excelencia Mar\'{\i}a de Maeztu, grant MDM-2017-0765 and Programa Severo Ochoa del Principado de Asturias (Spain); the Chulalongkorn Academic into Its 2nd Century Project Advancement Project, and the National Science, Research and Innovation Fund via the Program Management Unit for Human Resources \& Institutional Development, Research and Innovation, grant B39G670016 (Thailand); the Kavli Foundation; the Nvidia Corporation; the SuperMicro Corporation; the Welch Foundation, contract C-1845; and the Weston Havens Foundation (USA).
\end{acknowledgments}

\bibliography{auto_generated}
\cleardoublepage \appendix\section{The CMS Collaboration \label{app:collab}}\begin{sloppypar}\hyphenpenalty=5000\widowpenalty=500\clubpenalty=5000
\cmsinstitute{Yerevan Physics Institute, Yerevan, Armenia}
{\tolerance=6000
A.~Hayrapetyan, A.~Tumasyan\cmsAuthorMark{1}\cmsorcid{0009-0000-0684-6742}
\par}
\cmsinstitute{Institut f\"{u}r Hochenergiephysik, Vienna, Austria}
{\tolerance=6000
W.~Adam\cmsorcid{0000-0001-9099-4341}, J.W.~Andrejkovic, L.~Benato\cmsorcid{0000-0001-5135-7489}, T.~Bergauer\cmsorcid{0000-0002-5786-0293}, S.~Chatterjee\cmsorcid{0000-0003-2660-0349}, K.~Damanakis\cmsorcid{0000-0001-5389-2872}, M.~Dragicevic\cmsorcid{0000-0003-1967-6783}, P.S.~Hussain\cmsorcid{0000-0002-4825-5278}, M.~Jeitler\cmsAuthorMark{2}\cmsorcid{0000-0002-5141-9560}, N.~Krammer\cmsorcid{0000-0002-0548-0985}, A.~Li\cmsorcid{0000-0002-4547-116X}, D.~Liko\cmsorcid{0000-0002-3380-473X}, I.~Mikulec\cmsorcid{0000-0003-0385-2746}, J.~Schieck\cmsAuthorMark{2}\cmsorcid{0000-0002-1058-8093}, R.~Sch\"{o}fbeck\cmsAuthorMark{2}\cmsorcid{0000-0002-2332-8784}, D.~Schwarz\cmsorcid{0000-0002-3821-7331}, M.~Sonawane\cmsorcid{0000-0003-0510-7010}, W.~Waltenberger\cmsorcid{0000-0002-6215-7228}, C.-E.~Wulz\cmsAuthorMark{2}\cmsorcid{0000-0001-9226-5812}
% ...
\par}
\vskip\cmsinstskip
\dag:~Deceased\\
$^{1}$Also at Yerevan State University, Yerevan, Armenia\\
$^{2}$Also at TU Wien, Vienna, Austria\\
% ...
\end{sloppypar}
\end{document}